\pdfoutput=1 
\documentclass[fp,twocolumn]{jpsj3}
\usepackage{dcolumn}
\usepackage{bm}
\usepackage[pdftex]{graphicx}
\begin{document}

\def\runtitle{Dynamical Charge Structure Factor of a 1D Ionic Hubbard Model in the Low-Energy Region}
\def\runauthor{Nobuya Maeshima}

\title{Dynamical Charge Structure Factor of a One-Dimensional Ionic Hubbard Model in the Low-Energy Region}
\author{
  Yuhei Komaki$^1$,  Yuma Iwase$^2$, Shogo Yanagimatsu$^2$, Yoshiyuki Muta$^2$,  Nobuya Maeshima$^{3,4}$\thanks{maeshima@ims.tsukuba.ac.jp}, and Ken-ichi Hino$^{4,3}$
}

\inst{
  $^1$Doctoral Program in Materials Science, Graduate School of Pure and Applied Sciences, University of Tsukuba, Tsukuba, Ibaraki 305-8573, Japan\\
  $^2$College of Engineering Sciences, University of Tsukuba, Tsukuba 305-8573, Japan \\  
  $^3$Center for Computational Sciences, University of Tsukuba, Tsukuba 305-8577, Japan \\  
  $^4$Division of Materials Science, University of Tsukuba, Tsukuba 305-8573, Japan 
}
\recdate{\today}

\abst{
  We present a numerical study of the charge dynamical structure factor $N(k,\omega)$ of a one-dimensional (1D) ionic Hubbard model in the Mott insulator (MI) phase.
  We show that the low-energy spectrum of $N(k,\omega)$ is expressed in terms of the spin operators for the spin degrees of freedom.
  Numerical results for the spin degrees of freedom, obtained by the Lanczos diagonalization method, 
  well reproduce the low-energy spectrum of $N(k,\omega)$ of the 1D ionic Hubbard model.
  In addition, we show that these spectral peaks probe the dispersion of the spin-singlet excitations of the system and are observed in the wide parameter region of the MI phase.
}


\sloppy
\maketitle

\section{Introduction}\label{sec1}

The ionic Hubbard model, a Hubbard model with an alternating single-electron potential, has been extensively studied from several points of view in condensed matter physics,
such as a quantum phase transition~\cite{ortiz,resta,fabrizio,torio,manmana,otsuka,legeza,kampf},
the existence of a non-trivial intermediate phase~\cite{ortiz,resta,fabrizio,torio},
superconductivity~\cite{kuroki,watanabe}, ferromagnetism~\cite{lin}, ferroelectric properties of transition-metal oxides~\cite{ishihara},
and the neutral-ionic transition of mixed-stack organic complexes~\cite{nagaosa1,nagaosa2}.

Ground state properties of the ionic Hubbard model at half-filling are mainly governed by the competition between the on-site Coulomb interaction $U$ and the alternating potential $\Delta$.
For $\Delta>>U$, the periodicity of the lattice makes the system to the band insulator (BI) phase while,
for $U>>\Delta$, the strong on-site interaction leads the system to the Mott insulator (MI) phase.
The existence of the intermediate phase between them has been established in one-dimensional (1D) systems~\cite{ortiz,resta,fabrizio,torio}.

Properties of excited states of this model are also governed by $U$ and $\Delta$.
In the BI phase, excited states are described by using the particle, hole, and their pair excitations while,
in the MI phase, there exist low-lying spin excitations and charge excitations with the higher energy of the order of $U$.
Focusing on 1D systems, continuous unitary transformation method is applied to study excited states of the BI phase~\cite{hafez1,hafez2,torbati}.
In the vicinity of the boundary between BI and MI, domain wall excitations are investigated for the 1D system related to
a mixed-stack organic complex TTF-CA~\cite{nagaosa2}.
In the MI phase, a theoretical work by Katsura {\it et al.} has analytically demonstrated that,
in the optical conductivity $\sigma(\omega)$, spin excitations have finite intensity ascribed to the finite $\Delta$~\cite{katsura}.
Some of the authors and a coworker also have numerically confirmed that there appear spectral peaks resulting from the spin excitations in $\sigma(\omega)$~\cite{yokoi}.

Here it should be noted that, in general, excitations with the wave number $k=0$ are probed by $\sigma(\omega)$ while
the excitations with $k\ne0$ are detected by the dynamical charge structure factor $N(k,\omega)$~\cite{stephan}.
Thus the preceding studies for the MI phase imply that the spin excitations with $k\ne0$ can be detected by $N(k,\omega)$.
In this work, we have studied low-energy properties of $N(k,\omega)$ of the 1D ionic Hubbard model in the MI phase.
We represent the charge disproportionation per site and $N(k,\omega)$ in terms of the spin operators of the spin degrees of freedom of this model,
and then, numerically calculate these quantities by using the Lanczos diagonalization method.
Obtained results of $N^s(k,\omega)$, the contribution to $N(k,\omega)$ from the spin degrees of freedom, well reproduce the low-energy component of $N(k,\omega)$.
In addition, spectral peaks of $N^s(k,\omega)$ are in good agreement to the dispersion of the two-spinon singlet excitations obtained by the Bethe ansatz for the uniform chain.
Thus the spin-singlet excitations can be probed by examining $N(k,\omega)$ in the ionic Hubbard model.
We also confirmed that these spectral peaks resulting from the spin excitations are observed in the wide parameter region of the MI phase.

\section{Theoretical Framework}

The Hamiltonian of the 1D ionic Hubbard model reads
\begin{eqnarray}
  {\cal H} &=&- t\sum_{i \sigma} ( c^\dagger_{i\sigma}c_{i+1\sigma} +  H.c.)+ U\sum_{i} n_{i\uparrow}n_{i\downarrow} \nonumber \\
  &+& \sum_{i} e_i (n_i-1),
 \label{eq_ionic}
\end{eqnarray}
where $c^{\dagger}_{i\sigma}$ ($c_{i\sigma}$) is the creation (annihilation) operator of an electron with spin $\sigma$ at site $i$, $n_{i\sigma}=c^{\dagger}_{i\sigma}c_{i\sigma}$, and $n_{i}=n_{i\uparrow}+n_{i\downarrow}$.
The parameter $t$ denotes the nearest-neighbor transfer integral and the on-site Coulomb interaction is represented by $U$.
The site-dependent potential at site $i$ is defined by
\begin{equation}
e_i= -(-1)^i \frac{\Delta}{2} + \phi_i,
\end{equation}
where $\Delta$ is the strength of the alternating potential and $\phi_i$ represents an external scalar field.
we focus on the case of the half-filling and thus the total number of electrons is equivalent to the system size $N$.

The dynamical charge structure factor of the 1D ionic Hubbard model is given by
\begin{equation}
  N(k,\omega) =  \frac{1}{N} \sum_{\alpha} |\langle \alpha|n_k|0 \rangle|^2 \delta(\omega - E_\alpha + E_0 ), \label{eq_nkw}
\end{equation}
where
\begin{equation}
  n_{k}=\sum_l e^{ -i k l} n_l
\end{equation}
and $k=2\pi n/N (n=0,1,\cdots, N-1)$ is the wave number.
We note that $n_l$ is replaced by the charge disproportionation from the unity $\delta n_l \equiv n_l-1$ in order to subtract the trivial contribution from the total number of electrons for $k=0$.

To elucidate the contribution to $N(k, \omega)$ from the spin degrees of freedom in the MI phase,
we introduce the effective Hamiltonian of the model~(\ref{eq_ionic}) following Ref.~\citen{katsura}.
The effective Hamiltonian ${\cal H}^s$ is represented by
\begin{equation}
  {\cal H}^s = \sum_{i} \frac{4 t^2}{U  - e_i +e_{i+1}} \left( \bm{S}_i \cdot \bm{S}_{i+1} - \frac{1}{4} \right), \label{eq_Heisen}
\end{equation}
where $\bm{S}_i$ is the spin operator at site $i$.
The charge disproportionation is derived as
\begin{eqnarray}
  \delta n^s_{l} &=& \frac{\partial {\cal H}^s }{\partial \phi_l}  \nonumber \\
  &=&\sum_{j= l \pm 1}  \frac{8t^2U(e_l - e_j) }{[ U^2 - (e_l - e_j)^2]^2}  \left( \bm{S}_l \cdot \bm{S}_{j} - \frac{1}{4} \right) 
\end{eqnarray}
and the superscript ''$s$'' of quantities in the following means the contribution from the spin degree of freedom.

For $\phi_l = 0$, ${\cal H}^{s}$ and $\delta n^s_l$ are simplified as
\begin{equation}
  {\cal H}^s = K\sum_{l} \bm{S}_l \cdot \bm{S}_{l+1} \label{eq_1DHeisen}
\end{equation}
and
\begin{equation}
  \delta n^s_l = (-1)^l \eta \sum_{j=l \pm 1} \left( \bm{S}_l \cdot \bm{S}_j - \frac{1}{4} \right),  \label{eq_charge}
\end{equation}
where
\begin{equation}
  K=\frac{4t^2U}{U^2-\Delta^2} \quad {\rm and } \quad \eta =  \frac{8 t^2 U \Delta}{[ U^2 - \Delta^2 ]^2 }. \label{eq_eta}
\end{equation}
By replacing $n_{k}$ in Eq.~(\ref{eq_nkw}) by $\delta n^s_{k}=\sum_l e^{ -i kl} \delta n^s_l$ we obtain the contribution
to the dynamical charge structure factor from the spin degrees of freedom and it is represented by $N^s(k,\omega)$.
It should be noted that when computing $N^s(k,\omega)$ the eigenstate $|\alpha\rangle$ and the eigenenergy $E_\alpha$ of the Hamiltonian~(\ref{eq_ionic})
are also replaced by their counterparts, $|\alpha\rangle^s$ and $E_\alpha^s$ of the Heisenberg model~(\ref{eq_Heisen}).

By using these results, $\delta n^s_k$ is derived as
\begin{equation}
  \delta n^s_k = \sum_l e^{-i(k+\pi)l} \eta \sum_{j=l \pm 1} \bm{S}_l \cdot \bm{S}_{j},  \label{eq_dnk}
\end{equation}
where the constant 1/4 is omitted because it has no physically-important contribution to $N^s(k,\omega)$ for $\omega>0$.
Equation~(\ref{eq_dnk}) suggests that the state $\delta n^s_k |0\rangle^s$ has the wave number $k+\pi$.
Then the $\alpha$-th eigenstate $|\alpha\rangle^s$ of the model~(\ref{eq_Heisen})
with the finite matrix element $\langle \alpha|^s \delta n^s_l |0\rangle^s$ also has the wave number $k+\pi$ within the Hilbert space of the model.
That is, $N^s(k,\omega)$ probes excited states of the spin system with the wave number $k+\pi$, and 
therefore we represent and plot the charge dynamical structure factor as a function of $k+\pi \equiv k'$ in the following.

In actual calculations, we use the Lanczos diagonalization to calculate eigenstates including the ground state and other quantities.
Here it should be also noted that we impose the periodic boundary condition.
As for the dynamical charge structure factor the $\delta$-function is replaced by the Lorentzian with finite broadening $\epsilon=0.01t$.

\section{Result}

\begin{figure}
  \begin{center}
    \includegraphics[width=7.0cm]{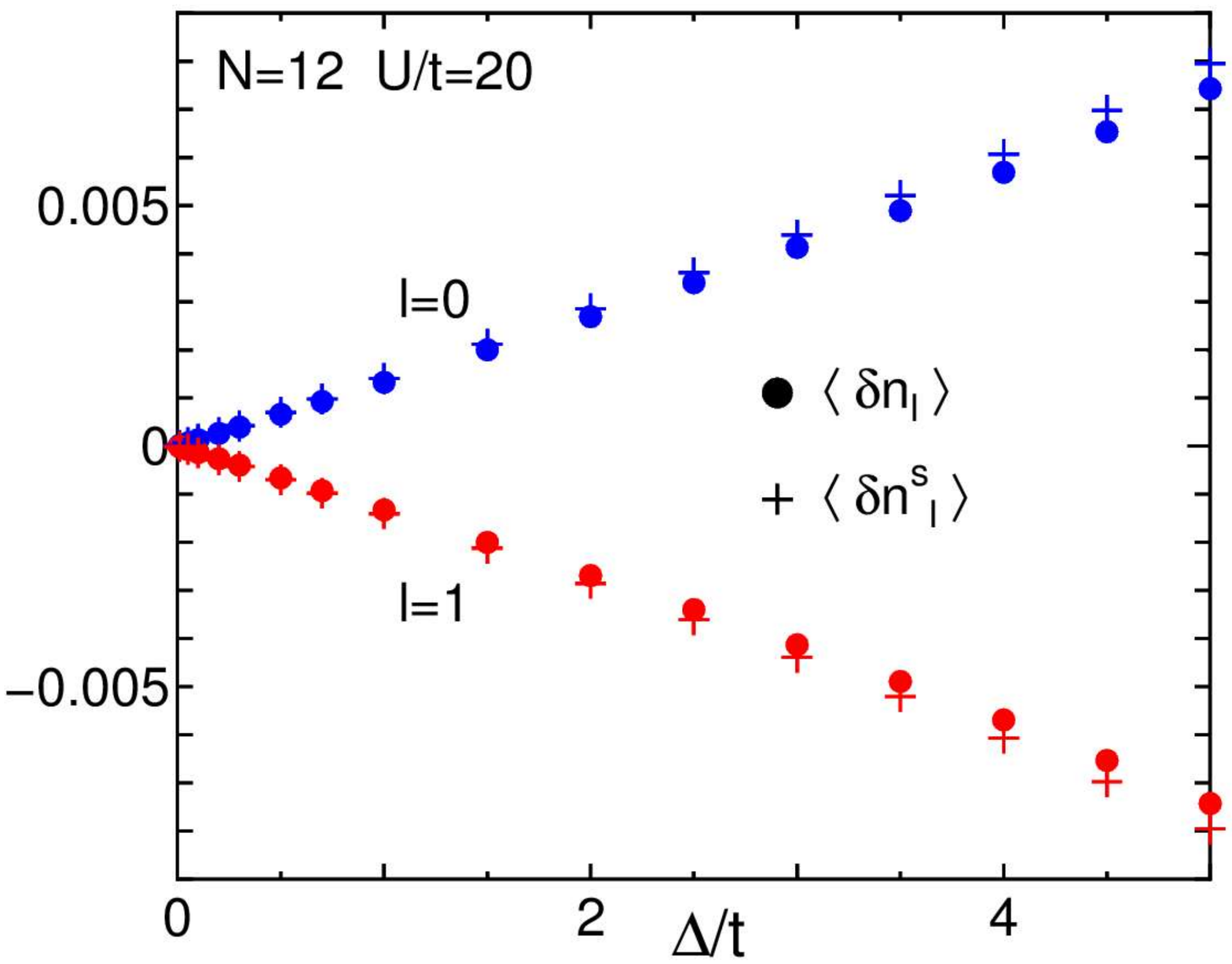}
  \caption{
    (Color online) The expectation values of $\delta n_l$ of the 1D ionic Hubbard model
    and its counterpart $\delta n^s_l$ of the corresponding 1D Heisenberg model.
    Here we set $U/t=20$ and $N=12$.
  }
  \label{fig_dn}  
  \end{center}
\end{figure}

Before analyzing the charge dynamical structure factor, we confirm that the charge disproportionation~(\ref{eq_charge}) gives valid results.
Figure~\ref{fig_dn} shows $\langle \delta n_l \rangle$ of the ionic Hubbard chain and $\langle \delta n^s_l \rangle$ of the Heisenberg chain
for even site ($l=0$) and odd site ($l=1$) in an $N=12$ system with $U/t=20$,
where $\langle O \rangle$ is the expectation value of an observable $O$ in the ground state.
It can be seen that the results of $\langle \delta n_l \rangle$ and $\langle \delta n^s_l \rangle$ are in good agreement, showing that the formula~(\ref{eq_charge}) works well.

\begin{figure}
  \begin{center}
    \includegraphics[width=7.0cm]{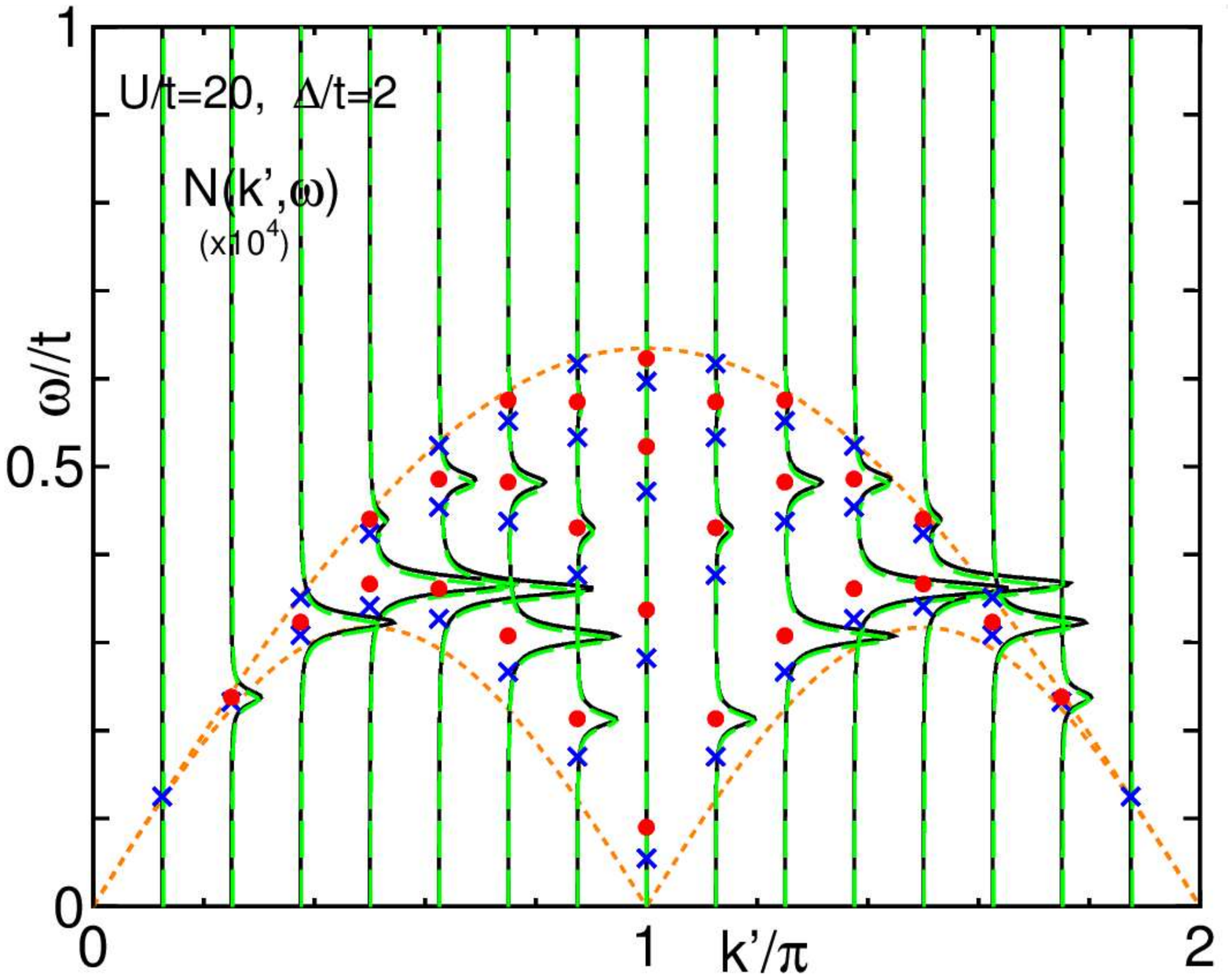}    
  \caption{
    (Color online) The charge dynamical structure factor $N(k,\omega)$ (green dotted lines) and $N^s(k,\omega)$ (black solid lines) of the system with $U/t=20$, $\Delta/t=2$, and $N=16$.
    The orange dotted lines represent the upper and lower bounds of the two-spinon excitations in the thermodynamic limit.    
    The red circles and the blue crosses show the two-spinon singlet excitations and the two-spinon triplet excitations for $N=16$, respectively.
  }
  \label{fig_nkw}  
  \end{center}
\end{figure}

Results of $N(k',\omega)$ and $N^s(k',\omega)$ for the system with $U/t=20$, $\Delta/t=2$, and $N=16$ are displayed in Fig.~\ref{fig_nkw}.
We can see that $N^s(k',\omega)$ shows almost the same peak structure as $N(k',\omega)$.
It is also found that the dominant spectral peaks are located around $k'=\pi/2$ and $3\pi/2$, which are clearly different from
those of the spin dynamical structure factor $S(k',\omega)$ with the dominant spectral peaks around $k'=\pi$~\cite{muller}.
However, the region where the peaks of $N(k',\omega)$ exist is almost the same as that of $S(k',\omega)$;
both are inside the upper and lower bounds of the two-spinon excitation in the thermodynamic limit~\cite{decpea,yamada},
suggesting that $N(k',\omega)$ mainly probes the two-spinon excitations~\cite{bougo,karbach1}.

To further examine excited states contributing to $N(k',\omega)$ we calculate the two-spinon singlet and triplet excitations of the
1D Heisenberg model with $N=16$ using the Bethe ansatz following Ref.~\citen{karbach2}.
It can be found that the main spectral peaks of $N(k',\omega)$ are coincident with the two-spinon singlet excitations
while those of $S(k',\omega)$ are coincident with the two-spinon triplet excitations~\cite{karbach2}.
Thus our results suggest that by probing $N(k',\omega)$ we obtain the dispersion relation of the two-spinon singlet excitations.
This is consistent with that $\delta n^s_l$ commutes with the total spin $\bm{S}_{\rm tot}=\sum_l \bm{S}_l$
and that $\sigma(\omega)$ of the 1D ionic Hubbard model detects the singlet excitations~\cite{katsura}.
Our results, of course, do not deny the possibility that $N(k',\omega)$ may have a finite contribution from the multi-spinon state
as $S(k',\omega)$ does~\cite{caux}. Investigating that point is, however, beyond the scope of this paper.

To examine the spectral intensity around $k'=\pi/2$ and $3\pi/2$, 
we generalize the formulae~(\ref{eq_charge}) and ~(\ref{eq_dnk}) derived from the Heisenberg Hamiltonian~(\ref{eq_Heisen}) to those of the XXZ model by
replacing $\bm{S}_i \cdot \bm{S}_j$ in these formulae by $(\bm{S}_i \cdot \bm{S}_j)_{\lambda} = S^x_i S^x_j + S^y_i S^y_j + \lambda S^z_i S^z_j$~\cite{katsura},
where $\lambda$ is the anisotropic parameter.
Further, by using the Jordan-Wigner transformation, the charge disproportionation~(\ref{eq_charge}) for the XXZ model is represented as 
\begin{eqnarray}
  \delta n^s_{k'}(\lambda) = \sum_l e^{-ik'l} \eta \sum_{j=l \pm 1} \left[ \frac{1}{2}\left(f^\dagger_l f_j + H.c. \right) \right. \nonumber \\
    +    \left. \lambda \left(n_l - \frac{1}{2} \right )\left(n_j - \frac{1}{2} \right ) \right],  \label{eq_dnk2}
\end{eqnarray}
where $f^{\dagger}_{l}$ ($f_{l}$) is the creation (annihilation) operator of the spinless fermion and $n_l=f^\dagger_l f_l$.
The charge dynamical structure factor $N^s(k',\omega; \lambda)$ for the XXZ model with $\lambda$ is also defined by 
the replacement of $n_{k}$ in Eq.~(\ref{eq_nkw}) by $\delta n^s_{k'} (\lambda)$ of Eq.~(\ref{eq_dnk2}).
Hence $N^s(k',\omega)$ is equal to $N^s(k',\omega; \lambda=1)$ in the following.

\begin{figure}
  \begin{center}
    \includegraphics[width=7.5cm]{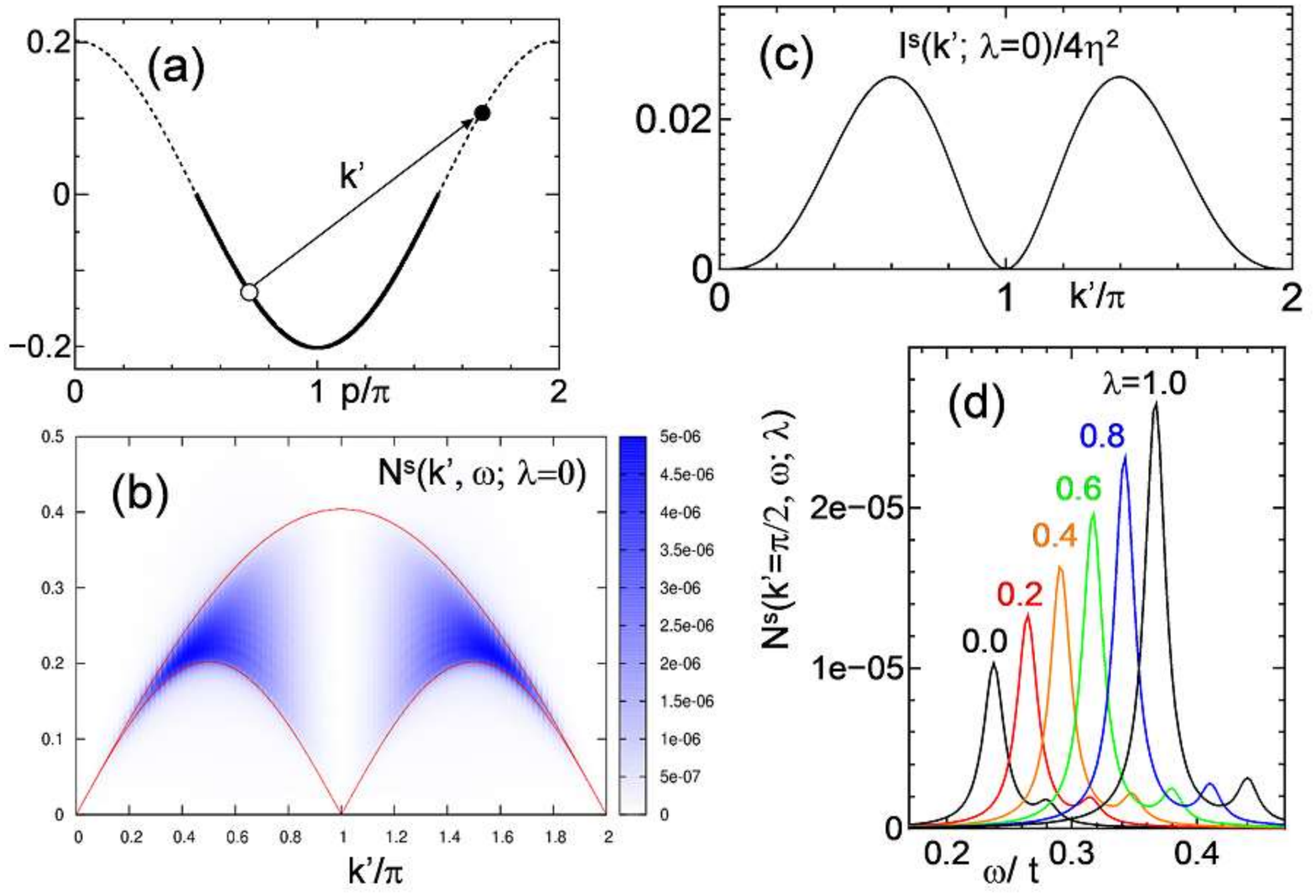}    
  \caption{
    (Color online) (a) The one-particle dispersion $e(p)=K \cos p$  (dotted line) of the XY model ($\lambda=0$) and an excitation process with the wave number $k'$ (arrow).
    The solid line for $\pi/2 \le p \ge 3\pi/2$ represents the Fermi sea representing the ground state of the XY model.
    (b) $N^s(k',\omega; \lambda=0)$ and the upper and lower bounds of the hole-particle pair excitation.
    (c) The integrated intensity $I^s(k'; \lambda=0)$. (d)  $N^s(k'=\pi/2,\omega; \lambda)$ for $N=16$.
    All the results are obtained for systems with $K/t=0.202020\cdots$, corresponding to $U/t=20$ and $\Delta/t=2$.
  }
  \label{fig_xxz}  
  \end{center}
\end{figure}

For $\lambda=0$, which corresponds to the XY model, $\delta n^s_{k'}(\lambda)$ is simplified as
\begin{eqnarray}
  \delta n^s_{k'}(\lambda=0) = -2\eta \sum_p f^\dagger_{p+k'} f_p \cos \left( \frac{k'}{2} \right) \nonumber \\
  \times \cos \left( \frac{k'}{2} + p\right), \label{eq_dnk3}
\end{eqnarray}
where $f_p = \sum_l e^{-i p l } f_l/\sqrt{N}$
is the annihilation operator of the spinless fermion with the wave number $p$ and $f^\dagger_p$ is its Hermite-conjugate.
Equation~(\ref{eq_dnk3}) tells us that $\delta n^s_{k'}(\lambda=0)$ consist of the process with the annihilation of the particle with the wave number $p$
and the creation of the particle with $p+k'$.
When operating $\delta n^s_{k'}(\lambda=0)$ to the ground state $|0\rangle^s$ of the XY model,
a typical key process corresponds to the excitation depicted by Fig.~\ref{fig_xxz} (a),
where the particle in the Fermi sea with the wave number $p\in [\pi/2, 3\pi/2]$ and the energy $e(p) = K \cos p$
is annihilated and the particle with $p+k'$ is created; this is considered to be the hole-particle pair excitation process.

The charge dynamical structure factor for $\lambda=0$ is obtained as
\begin{eqnarray}
  N^s(k',\omega; \lambda=0) =  \frac{4\eta^2}{N} \sum_{\pi/2 \le p \le 3\pi/2 } 
  \cos^2 \left( \frac{k'}{2} \right)   \nonumber \\
  \times \cos^2 \left( \frac{k'}{2} + p\right)
  \Theta \left[ e(p+k') \right] \Theta \left[ -e(p) \right]   \nonumber \\
  \times \delta[ \omega - e(p+k') + e(p) ], \label{eq_nkw4}
\end{eqnarray}
where $\Theta [ x ]$ is the step function with $\Theta[x]=1(0)$ for $x \ge 0 (<0)$.
We numerically evaluate Eq.~(\ref{eq_nkw4}) with replacing the $\delta$-function by the Lorentzian with finite broadening $\epsilon=0.01t$.
The obtained result shown in Fig.~\ref{fig_xxz} (b) is qualitatively similar to that of the Heisenberg model ($\lambda=1$) shown in Fig.~\ref{fig_nkw}.
The formula~(\ref{eq_nkw4}) also tells us that the coefficient $ \cos^2 \left( k'/2 \right)$ gives the vanishing spectral intensity around $k'\sim \pi$.
In addition, the spectral intensity around $k'\sim 0$ is also suppressed because the associated excitations occur only for $p\sim \pi/2$ or $3\pi/2$
and then the coefficient $\cos^2 (k'/2 + p)$ has a small value. Hence the finite spectral intensity is observed in the remaining areas.
The integrated intensity
\begin{equation}
  I^s(k'; \lambda) = \int_0^\infty d\omega N^s(k',\omega; \lambda)
\end{equation}
for $\lambda=0$, plotted in Fig.~\ref{fig_xxz} (c), shows that the peaks spread over the wide range centered on $k' \sim 0.6\pi$ or $1.4\pi/$,
which suggests that the excitations with $k'=\pi/2$ or $3\pi/2$ are not the extremely dominant excitation processes.
As $\lambda$ increases, the spectrum does not change qualitatively while the lowest spectral peak develops [see Fig.~\ref{fig_xxz} (c)];
this is possibly attributed to the excitonic effect induced by finite $\lambda$~\cite{katsura}.
The blue shift of the peaks is merely caused by the increasing bandwidth of the two-spinon excitations.

\begin{figure}
  \begin{center}
    \includegraphics[width=7.0cm]{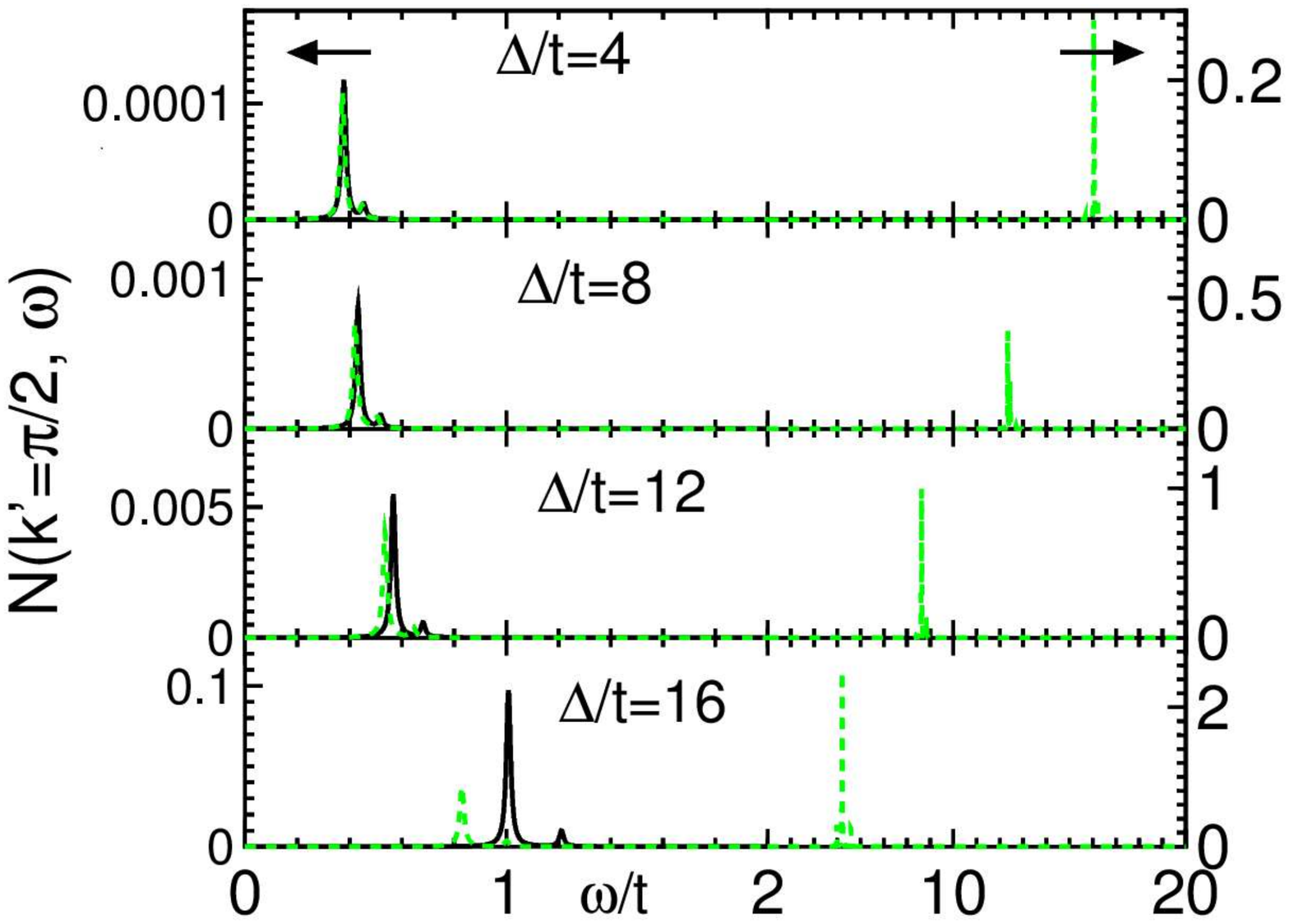}    
  \caption{
    (Color online) $N(k'=\pi/2,\omega)$ (green dotted lines) and $N^s(k'=\pi/2,\omega)$ (black solid lines) of the system with $U/t=20$ and $N=16$.
  }
  \label{fig_nkw_nk4}  
  \end{center}
\end{figure}

\begin{figure}
  \begin{center}
    \includegraphics[width=7.0cm]{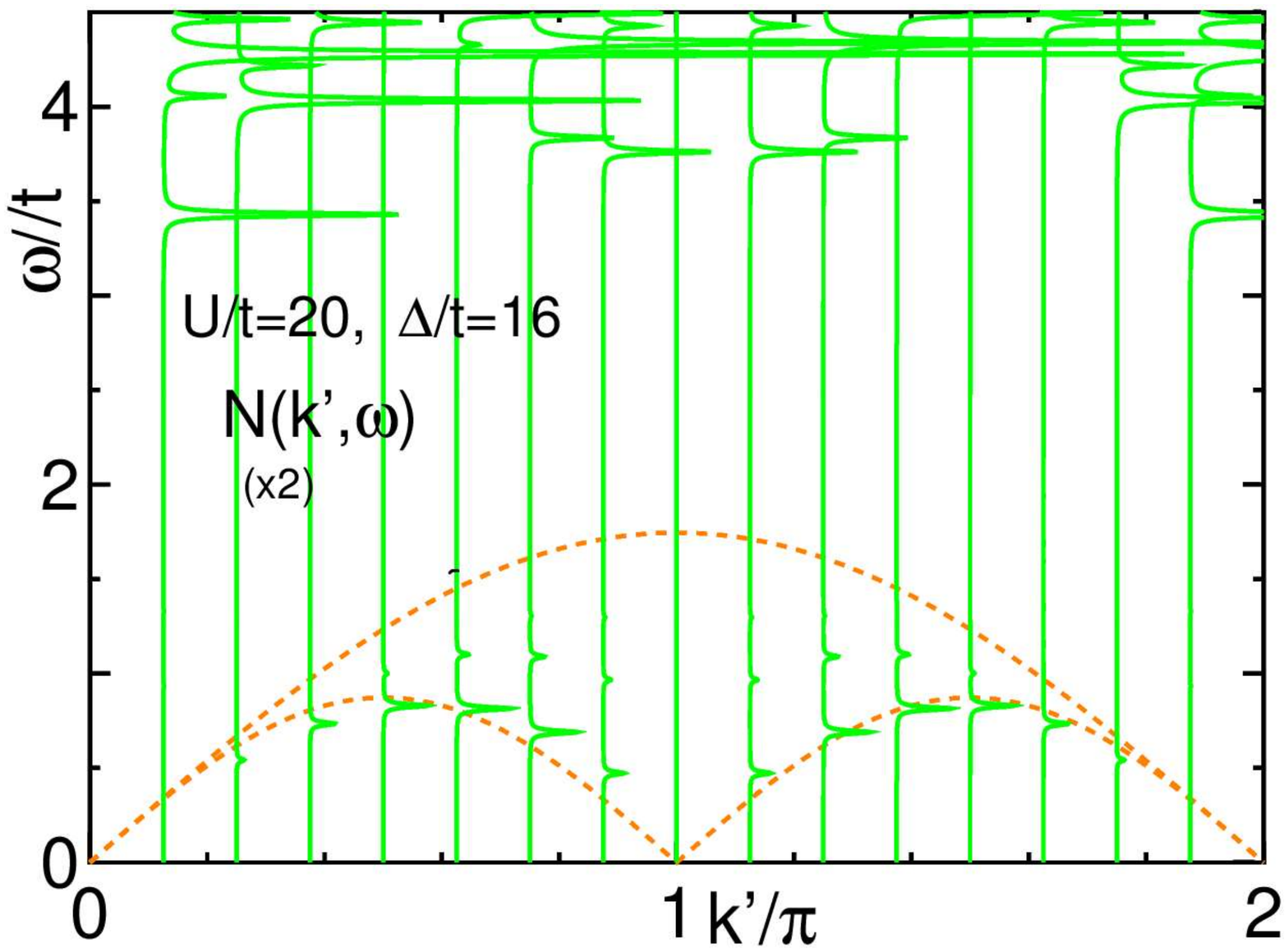}    
  \caption{
    (Color online) $N(k',\omega)$ (green solid lines) of the system with $U/t=20$, $\Delta/t=16$, and $N=16$.
    The orange dotted lines are the upper and the lower bounds of the two-spinon continuum.
  }
  \label{fig_nkw_U20D16}  
  \end{center}
\end{figure}

All the results shown above are obtained for systems with sufficiently large $U$ and small $\Delta$, where the perturbation theory can be undoubtedly applied.
To clarify the region where the perturbative treatment works well,
we calculate $N^s(k',\omega)$ and $N(k',\omega)$ at $k'=\pi/2$ for larger $\Delta/t$ (see Fig.~\ref{fig_nkw_nk4}).
It can be found that $N^s(k',\omega)$ quantitatively reproduce the low-energy spectral peaks of $N(k',\omega)$ up to around $\Delta/t = 12$.
For $\Delta/t=16$, we can observe apparent deviation between them;
the lowest excitation gap of $N(k',\omega)$ is slightly smaller than that of $N^s(k',\omega)$
and the peak intensity of $N(k',\omega)$ is also smaller than that of $N^s(k',\omega)$.
The low-lying peaks for $\Delta/t=16$ are, however, well separated from the charge transfer (CT) excitations with the CT gap $\Delta_{\rm CT} \sim 4t$
and are roughly inside the upper and the lower bounds of the two-spinon continuum of the Heisenberg model, as shown in Fig.~\ref{fig_nkw_U20D16}.
Thus the significant part of the low-lying excited states are possibly caused by the spin excitation even for $\Delta/t=16$,
some part of which would be from other excitations including the domain wall excitations~\cite{nagaosa2}.

\begin{figure}
  \begin{center}
    \includegraphics[width=7.0cm]{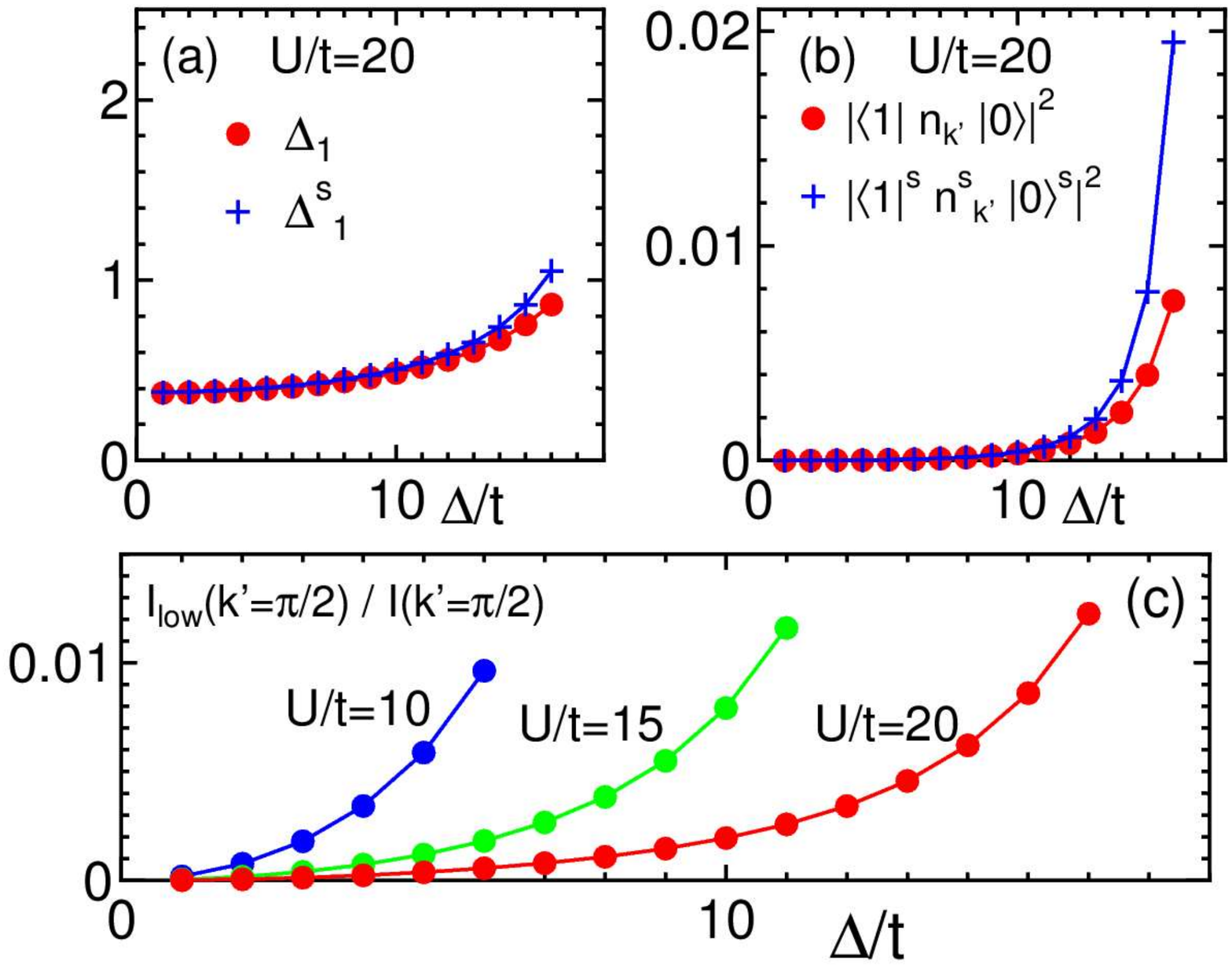}    
  \caption{
    (Color online) (a) The lowest excitation gap $\Delta_1 = E_1 -E_0$ for $N(k'=\pi/2,\omega)$ of the 1D ionic Hubbard model (red circles)
    and its counterpart $\Delta^s_1 = E^s_1 -E^s_0$ of the Heisenberg model (blue crosses).
    (b) The matrix element $|\langle 1 |n_{k'} |0\rangle|^2 $ (red circles) and $|\langle 1 |^s n^s_{k'} |0\rangle^s|^2 $ (blue crosses) at $k'=\pi/2$.
    (c) The ratio of the integrated intensity $I_{\rm low}(k'=\pi/2)/I(k'=\pi/2)$.
    All the calculations are carried out for systems with $N=12$. 
  }
  \label{fig_intnk}
  \end{center}
\end{figure}

As discussed above, the low-lying spectral peaks of $N(k',\omega)$ are basically attributed to the two-spinon singlet excitations in the wide region of the MI phase.
In particular, the excitation energies of these peaks are well explained by using the Heisenberg model.
To examine this point, we plot the lowest excitation energy $\Delta_1 = E_1 -E_0$ for $N(k'=\pi/2,\omega)$ of the 1D ionic Hubbard model in Fig.~\ref{fig_intnk} (a),
where $\Delta_1$ is well coincident with its counterpart $\Delta^s_1 = E^s_1 -E^s_0$  of the Heisenberg model; even at $\Delta/t=16$, $\Delta_1$ is only about 20\% smaller than $\Delta^s_1$.
By contrast, the corresponding matrix element $|\langle 1 |n_{k'} |0\rangle|^2 $ of the 1D ionic Hubbard model deviates from 
$|\langle 1 |^s n^s_{k'} |0\rangle^s|^2 $ of the Heisenberg model for large $\Delta/t$.
At $\Delta/t=16$, $|\langle 1 |n_{k'} |0\rangle|^2 $ is roughly half of $|\langle 1 |^s n^s_{k'} |0\rangle^s|^2 $.
Thus we conclude that the energy dispersion of the two-spinon singlet excitations can be detected with some degree of accuracy even for considerably large $\Delta/t$.

It is well-known that the charge dynamical structure factor $N(k',\omega)$ can be experimentally observed~\cite{EELS}.
From the theoretical point of view, the key ingredient to detect the contribution to $N(k',\omega)$ from the spin degrees of freedom is its total spectral intensity
in addition to that the spin excitations are sufficiently separated from the CT excitations.
To discuss this point, we calculated the total intensity
\begin{equation}
  I(k') = \int_0^\infty d\omega N(k',\omega)
\end{equation}
of the 1D ionic Hubbard model and its low-energy contribution
\begin{equation}
  I_{\rm low} (k') = \int_0^{\omega_u} d\omega N^s(k',\omega),
\end{equation}
where the upper limit $\omega_u$ is set to just below $\Delta_{\rm CT}$ for each $k'$.
Figure~\ref{fig_intnk} (c) shows the ratio $I_{\rm low}(k')/I(k')$ for $U/t=10,15$ and 20 at $k'=\pi/2$.
Calculations are carried out in the parameter region where the low-lying peaks are separated from the CT peaks for each $U/t$.
It can be found that the ratio monotonically increases and reaches at most the order of $10^{-2}$ within the MI phase where the spin excitations are sufficiently separated from the CT excitations. This increasing behavior results from the $\Delta$-dependence of $\eta$ in Eq.~(\ref{eq_eta}).
In the vicinity of the phase boundary between the MI phase and the BI phase, much larger value of  $I_{\rm low}(k')/I_{\rm tot}(k')$ is expected.
Since the CT excitations approach the spin excitations in that region, however, it is difficult to distinguish between them.
Therefore, fine tuning of physical parameters including $\Delta/t$ in the MI phase would be necessary for observing $N^s(k',\omega)$.

\section{Summary}

In this work, we have studied the low-energy spectrum of the charge dynamical structure factor $N(k,\omega)$ of the 1D ionic Hubbard model in the MI phase.
We represent the charge disproportionation $\delta n_l$ and $N(k,\omega)$ in terms of the spin operators of the 1D Heisenberg model and 
numerically calculated these quantities by using the Lanczos diagonalization.
Then the notable point is that $N^s(k,\omega)$, the contribution from the spin degrees of freedom, probes excited states of the Heisenberg model with the wave number $k+\pi=k'$.
Obtained results of $N^s(k',\omega)$ well reproduce the low-energy component of $N(k',\omega)$, and 
spectral peaks of $N^s(k',\omega)$ are in good agreement to the dispersion of the two-spinon singlet excitations obtained by the Bethe ansatz for the 1D Heisenberg model.
Most of the large peaks are located around $k' \sim \pi/2 $ or $3\pi/2$, which is qualitatively explained on the basis of the analysis for the 1D XXZ model.
We confirmed that the low-lying spectral peaks of $N(k',\omega)$ resulting from the spin excitations are observed in the wide parameter region of the MI phase.
We have also shown that, although the low-energy spectral intensity is basically small, it rapidly increases with $\Delta$ owing to increasing $\eta$.
Therefore, in order to experimentally observe $N^s(k',\omega)$, it would be necessary to use the system where
the parameter $\Delta/t$ is large but the spin excitations can be distinguished from the CT excitations.

This work was supported by JSPS KAKENHI Grant Number 26800163 from the Ministry of Education, Culture, Sports, Science and Technology (MEXT), Japan.


\end{document}